\documentclass[sigconf]{acmart}

\usepackage{graphicx}
\usepackage{subcaption}
\usepackage{amsmath}
\usepackage{amsfonts}
\usepackage{booktabs}
\copyrightyear{2025}
\acmYear{2025}
\setcopyright{cc}
\setcctype{by}
\acmConference[FAccT '25]{The 2025 ACM Conference on Fairness,
Accountability, and Transparency}{June 23--26, 2025}{Athens, Greece}
\acmBooktitle{The 2025 ACM Conference on Fairness, Accountability, and Transparency (FAccT '25), June 23--26, 2025, Athens,
Greece}
\acmDOI{10.1145/3715275.3732099}\acmISBN{979-8-4007-1482-5/2025/06}

\begin{document}

\title{Not Even Nice Work If You Can Get It; A Longitudinal Study of Uber's Algorithmic Pay and Pricing}

\author{Reuben Binns}
\email{firstname.lastname@cs.ox.ac.uk}
%

\affiliation{%
  \institution{Department of Computer Science, University of Oxford}
  \city{Oxford}
  \country{United Kingdom}
}

\authornote{Both authors contributed equally to the paper}

\author{Jake Stein}
\email{firstname.lastname@cs.ox.ac.uk}
%

\affiliation{%
  \institution{Department of Computer Science, University of Oxford}
  \city{Oxford}
  \country{United Kingdom}
}

\authornotemark[1]

\author{Siddhartha Datta}
\email{firstname.lastname@cs.ox.ac.uk}
%

\affiliation{%
  \institution{Department of Computer Science, University of Oxford}
  \city{Oxford}
  \country{United Kingdom}
}

\author{Max Van Kleek}
\email{firstname.prefix.lastname@cs.ox.ac.uk}
%

\affiliation{%
  \institution{Department of Computer Science, University of Oxford}
  \city{Oxford}
  \country{United Kingdom}
}

\author{Nigel Shadbolt}
\email{firstname.lastname@cs.ox.ac.uk}
%

\affiliation{%
  \institution{Department of Computer Science, University of Oxford}
  \city{Oxford}
  \country{United Kingdom}
}


\renewcommand{\shortauthors}{}

\begin{abstract}
Ride-sharing platforms like Uber market themselves as enabling `flexibility' for their workforce, meaning that drivers are expected to anticipate when and where the algorithm will allocate them jobs, and how well remunerated those jobs will be. In this work we describe our process of participatory action research with drivers and trade union organisers, culminating in a participatory audit of Uber's algorithmic pay and work allocation, before and after the introduction of dynamic pricing. Through longitudinal analysis of 1.5 million trips from 258 drivers in the UK, we find that after dynamic pricing, pay has decreased, Uber's cut has increased, job allocation and pay is less predictable, inequality between drivers is increased, and drivers spend more time waiting for jobs. In addition to these findings, we provide methodological and theoretical contributions to algorithm auditing, gig work, and the emerging practice of worker data science.

\end{abstract}

\maketitle

\section{Introduction}



Uber is an on-demand ride-hailing service which connects passengers with drivers, and sets both passenger prices and driver pay. Until a 2021 Supreme Court case, Uber drivers in the UK were unlawfully misclassified as independent contractors rather than workers, and for the purposes of minimum wage law, were only considered `working' while carrying passengers from pickup to dropoff. The Court upheld the Employment Tribunal's view that working time also includes whenever a driver is logged into the app and `available to accept a trip request'\cite{adams2022uber}. While Uber has implemented some changes in partial compliance with the Court judgement, they still only pay drivers for time on trip and en route to pickup. This means Uber can increase availability of rides without increasing their costs, as they can grow the pool of drivers waiting around for trips without pay. A key conflict therefore remains between Uber's restricted definition of working time, and a more expansive definition which includes all the additional time in between jobs.

When Uber first launched in the UK, prices were determined by the distance and duration of the trip, with Uber taking a fixed percentage of the passenger's fare, first 20\% and then later raised to 25\%. In Q1 of 2023, Uber introduced `dynamic' pricing in London, which means fares are no longer a simple function of time and distance. Instead the price the passenger pays and the fee the driver receives vary independently of each other, and are calculated dynamically based on location of pick up and drop off, time of day / week / year, probability of driver / passenger cancellation, and other factors undisclosed by Uber. This means that Uber's take rate (the percentage they keep) is no longer stable but varies trip-by-trip. Dynamic pricing has been opposed by drivers on the grounds that it creates opacity and uncertainty around how pay is calculated, what cut Uber may take, and how much a driver can expect to make in a given hour or location \cite{wired_drivers}.

The amount that a passenger pays for a trip is not available to drivers in the app, and Uber bars drivers from asking their passengers for this information directly, so it is not possible for drivers to work out Uber's cut on any given journey. In December 2024, Uber began disclosing a weekly average take rate to drivers in the app, but this still does not include a trip-by-trip breakdown. However, some media reports suggest Uber's cut can be as high as 47\% \cite{bbc_impossible}. An Uber spokesman responded that such take rates are `not the case when averaged out over the week', and the take rate `remained stable for many years'. However, without both trip-level and aggregate analysis over time and across multiple drivers, the impacts of dynamic pricing are hard to assess.

In this work, we present an audit of Uber's pay and work allocation algorithms, to understand: how much Uber drivers are being paid for their work (contrasting the Employment Tribunal's definition of working time with Uber's own); how dynamic pricing has impacted the cut that Uber takes from the rider fare, and how this distributed; and related metrics like time spent waiting for jobs, and changes in the predictability of work opportunities. We partnered with Worker Info Exchange (WIE), a non-profit organisation representing gig workers in the UK and Europe. WIE helps gig workers access and gain insight from data collected from them at work, and has empowered hundreds of workers to use their rights under data protection law to access a copy of all their data from Uber, including detailed trip data.
258 drivers volunteered their data covering over 1.5 million trips between 2016-2024. We engaged with WIE and drivers in participatory action research over a period of two years, mainly on-site in London, UK. Our study design, methods, research questions, and findings have emerged through close collaboration with WIE and individual gig workers. 

The primary aims of the audit, developed in collaboration with WIE and drivers, were to measure the following:
    \begin{itemize}
        \item \textbf{RQ1:Pay}: How much are drivers paid per hour worked (given contrasting definitions of working time)?
        \item \textbf{RQ2:Take Rate}: How are fares split between Uber vs drivers?
        \item \textbf{RQ3:Utilisation}: How much working time is spent on trips vs on standby?
        \item \textbf{RQ4:Predictability}: Has the predictability of work opportunities changed?
    \end{itemize}

Our longitudinal data allows us to assess how these key aspects of work and pay are changing over time. In the case of take rates, we can see how dynamic pricing has impacted what percentage of a passenger's fare goes to the driver and how much is kept by Uber. We find that in the year following dynamic pricing, Uber took a larger median average cut, with take rates as high as 50\% on some trips. In real terms, average pay per hour has been roughly stagnant since 2016, and is lower in the year following the introduction of dynamic pricing, even on Uber's own (limited) definition of working time. While a small number of drivers are earning more per hour worked, the majority are earning less. Post-dynamic pricing, Uber's passengers now pay higher prices, but drivers are not better off. Utilisation has decreased, meaning drivers spend more time on unpaid `standby' while waiting to be allocated a job. Finally, work is less predictable after the introduction of dynamic pricing, meaning how much a driver can expect to earn from a trip of a certain duration / distance, at a given time, can no longer be predicted based on similar trips from before dynamic pricing.

In addition to reporting findings from our algorithm audit, we provide methodological and theoretical contributions to existing literature on algorithm auditing, gig work, and the emerging practice of `participatory worker data science'\cite{calacci2022bargaining,gallagher2023digital}. Methodologically, we demonstrate an alternative approach to algorithm audits which uses large-scale data subject access requests (DSARs). Unlike other approaches to external audits which often rely on black box observations or crowdsourced screenshots, this data derives directly from the algorithm deployer’s own back-end, revealing schemas, attributes, identifiers, profiles and scores applied to data subjects. To our knowledge, this is the first contribution to the algorithm auditing field of a large-scale audit based on DSARs.

\section{Background}

\subsection{Algorithmic Management and Worker Resistance}
Algorithmic, data driven management of human workers on gig platforms like Uber has been studied extensively in a variety of research domains. Lee et al. studied drivers' use of online forums to make sense of work allocation algorithms \cite{lee2015working}. Compared to human managers, workers perceive algorithmic management decisions to be less fair and trustworthy, \cite{lee2018understanding}. Further work in this vein has used participatory design to co-design interventions with workers, including: to understand their needs for well-being \cite{zhang2022algorithmic}; to elicit stakeholder preferences for addressing worker well-being \cite{hsieh2023co}; and to explore `data probes' — interactive visualizations of workers’ data - as training tools for policymakers \cite{zhang2024data}. Ma et al. study how gig work platform designers sense of morality regarding the plight of workers affects their own practice \cite{ma2024my}. Through a study of mobility platforms in Jakarta, Qadri and D'Ignazio chart how workers develop their own `ways of seeing' urban mobility markets, which can both contest and complement the algorithmic visions of the platform \cite{qadri2022seeing}.




Recent work has explored various tools which could be deployed by and for workers themselves. Calacci et al., designed and deployed Shipt Calculator, a tool developed in collaboration with non-profit worker groups to track and share aggregate data about worker pay, increasing wage transparency \cite{calacci2022bargaining} (this provides direct inspiration for parts of our audit below).  Do et al. explore a variety of `sousveillance' tools to allow workers to monitor their algorithmic managers \cite{do2024designing}. Similarly, Gallagher et al. develop `participatory worker data science' \cite{gallagher2023digital}, through the `Workers' Observatory', a collaboration between workers and researchers. In previous work, also in collaboration with WIE, we explored possible data governance and infrastructure to enable bottom-up data institutions for gig workers \cite{stein2023you}.

Some prior studies involve auditing of Uber's pricing algorithm. These include a 2015 study of Uber's surge pricing algorithm in major US cities \cite{chen2015peeking}; a study of passenger pricing which found disparate impacts in pricing between neighbourhoods \cite{pandey2021disparate}; and Brodeur and Nield's 2018 analysis of weather shocks on mobility gig work platforms, which demonstrates how dynamic pricing can reduce driver income on Uber (among other platforms) \cite{brodeur2018empirical}. Most recently, the FairFare project crowdsources and analyzes workers’ data to estimate Uber's take rate in the US\cite{calacci2025fairfare}.

Despite confirmation of their legal status as workers in the UK following the 2021 Supreme Court judgement, Uber continues to inculcate the ideal of drivers as entrepreneurs. Prior work has argued that the ideological construction of the driver as an atomised individual, an `entrepreneurial self' \cite{helene2024end, purcell2022least}, is an important aspect of Uber's legitimization in the eyes of policymakers, regulators, and drivers themselves. Gregory \& Sadowski argue that gig workers are encouraged to develop their own human capital in perverse ways to make themselves `more productive, more desirable workers'\cite{gregory2021biopolitical}.




While worker resistance and solidarity does persist\cite{tirapani2023revisiting}, such efforts are often made harder under algorithmic management, which replaces human supervisors `with a set of highly depersonalised and disembodied commands' \cite{walker2021you}. Rosenblat and Stark argue that Uber relies on `information and power asymmetries' and `rhetorical invocations of digital technology and algorithms' to control workers and structure asymmetric corporate relationships in their favour \cite{rosenblat2016algorithmic}. Many workers buy into the idea of themselves as entrepreneurs, which aims merely to improve individual conditions without challenging the platforms' business model \cite{barratt2020m}. Nevertheless, surveys of platform workers in the UK suggest support for more labour rights, representation, and voice is high \cite{martindale2024platform}.  Finally, not all gig work platforms treat their workers the same. Kusk et al conducted an ethnographic study of workers on the Wolt food delivery app \cite{kusk2022working}, finding that its practices of algorithmic management are perceived as less `harsh' and `despotic' than those of Uber. Alternative ownership structures, with better working conditions, are also imaginable\cite{scholz2016platform,james2021gig}.





\subsection{Algorithm Auditing with Data Subject Access Requests}

The nascent sub-field of algorithm auditing has developed rapidly over the last decade. Audits can be performed internally by the organisation deploying the algorithm, or externally by a third party, either consensually (i.e. upon invitation by the deployer) or non-consensually (e.g. by researchers and activists without permission from the deployer) \cite{raji2023change}; our audit is external and non-consensual. While many organisations are voluntarily engaging in audits \cite{wilson2021building}, external audits remain an important tool to uncover algorithmic harms and seek redress for affected communities where internal audits and harm mitigation are not forthcoming. External audits have covered topics including: racial discrimination in online ad delivery \cite{sweeney2013discrimination}, job ads \cite{imana2021auditing}, personalisation in recommender systems such as Youtube \cite{hussein2020measuring}, and gender and racial bias in facial recognition technology \cite{buolamwini2018gender}. However, such external audits can be limited by their reliance on system behaviour that is publicly observable to researchers from the outside. They do not have access to the internal, back-end systems of the algorithm deployer. As such, researchers are often in the dark about exactly what data is collected, and how users are scored or classified on the basis of that data. Privacy concerns mean individual-level data is also not easily accessible, and even where scrapable, its unconsented use is at odds with prevailing research ethics norms. Such issues can significantly affect the kinds of questions that can be asked and the ecological validity of external audits.

However, a complementary data source for algorithm audits is also available: legally mandated data access rights such as Data Subject Access Requests (DSARs). Research participants in the UK and EU can access their data from organisations using a DSAR under Article 15 of the General Data Protection Regulation, as well as a subset of portable machine readable data under Article 20. DSARs have already been used by researchers and data rights activists to investigate organisations’ data practices \cite{ausloos2020researching}, including those of complex tech infrastructures and data pipelines such as Apple’s Siri voice assistant \cite{veale2018data}.

So far, such research has mainly sought understanding with relatively small samples. But with larger samples, it may be possible to investigate systemic effects. Recent proposals include open-source data donation frameworks (e.g. OSD2F ) which combine multiple research participants DSARs while preserving their privacy \cite{boeschoten2022framework}. A similar approach has been pursued for data obtained under Article 20 \cite{zwiebelmann2021data}, and for data download functions (e.g. to study ad targeting on X/Twitter \cite{wei2020twitter}). Data rights activists have also employed DSARs for studying algorithms; notable examples include Open SCHUFA\footnote{\url{openschufa.de}}, an attempt to reveal the biases of a ubiquitous German credit scoring algorithm, and Hestia Labs, which provides DSAR tools for collectives, including Uber drivers.\footnote{\url{hestialabs.org}} Our approach is inspired by these efforts.

\section{Methodology, Data Source, \& Participatory Audit Design}

In this section, we describe our methodology, data source, and approach to participatory audit design.

\subsection{Methodology}
Our approach to this work was informed by previously identified challenges and limitations associated with algorithm auditing and the use of data rights in worker justice. First, we aimed for our research and auditing efforts to be aligned to and embedded within existing struggles for worker justice (following  \cite{dencik2024data}). We drew from Katell, Young et al.’s framework for developing `situated interventions for algorithmic equity' \cite{katell2020toward}, which emphasises first working with organisers to document the degree of current efforts, resources available and setting milestones or conditions for success. We also followed a participatory action research (PAR) methodology, meaning that we iteratively identified problems, co-designed solutions, and reflected on the efficacy of both the solution produced and the design process itself. This was made possible by our partnership with WIE, which involved two researchers regularly spending days embedded in their workplace, enabling us to not only work on the specifics of the audit, but understand the issues and needs of organisers and drivers on a day-to-day level, hear about their latest campaigns, and attend trade union meetings and protests. All of these activities helped us to understand ongoing issues and actions, and to gain feedback on our ongoing auditing work.

As previous work has argued, individual-level data received in subject access requests may not enable the study of collective impacts, and its collection and schematisation will likely reflect the interests of Uber, rather than its workers\cite{calacci2022bargaining}. We sought to overcome these shortcomings of individual-level data access, by pursuing the `collective leveraging of data rights' \cite{stein2022workers}, while also working closely with organisers and workers to critically interrogate Uber's schemas and metrics, and attempt to re-construct alternative metrics from it which are more conducive to their interests.

As Birhane et al caution \cite{birhane2024ai}, AI audits often fall short due to the following limitations: focusing only on evaluation, without considering accountability; failing to engage with excluded stakeholders or foster collective action; or failing to use the full range of audit methodologies, focusing narrowly on popular metrics (e.g. `group fairness'). We aimed to overcome such limitations through our approach. First, the audit design is already informed by WIE's day-to-day work, which focuses on worker organising, policy advocacy, strategic litigation, and other work aimed at holding platforms to account. WIE's close links with affected stakeholders - in this case, Uber drivers themselves - allowed us to directly involve them in the design of the audit and formulation of research questions. Driver participation was essential, not only in making the data collection possible but also in motivating and shaping the analysis. Finally, we took an open-ended approach to audit metrics, creating a bespoke set of analyses to highlight insights of concern to the stakeholders, rather than attempting to apply existing formal metrics (e.g. from the algorithmic fairness literature).


\subsection{Data Source}
\label{subsection:datasource}
This section briefly describes the primary data source - a collection of 258 DSAR responses - including their provenance and governance. Individual drivers sign an affidavit with WIE to allow them to make and receive DSARs on their behalf, for the purposes of worker advocacy and research. We created a data transfer agreement, undertook a data protection impact assessment, and applied and received ethical approval from our institution's Research ethics committee, under reference `CS C1A 23 016'.

When WIE receives data from gig work platforms, it takes the shape of many disorganized files that are not readily understandable without substantial interpretation, preprocessing, and cleaning. In the case of Uber driver DSARs, we typically found between 40-45 different files, of which around 35 were CSVs containing a wide variety of data from different parts of Uber's backend systems. We worked on-site initially, to understand the varying data schemas and identify fields that would be likely required for further analysis. We created an automated tool to allow staff who lacked requisite technical skills to process the data into more interpretable forms themselves. This involved cleaning, reformatting, converting and joining data from multiple tables (see \ref{fig:DSAR} in Appendix). Given the sensitivity of the data, this tool ran locally on a dedicated laptop located in WIE's offices accessible only to a senior member of staff. For the audit itself, where we as researchers required access to larger-scale aggregate data on our own infrastructure, we created an aggregation pipeline which pseudonymised or stripped entirely a range of sensitive fields. Direct identifiers such as name, email, etc, were removed, and the platform-assigned driver ID was hashed, enabling the driver themself or WIE (but not us as researchers) to re-identify the driver if required.





\subsection{Participatory Worker Data Science}

In order to facilitate worker involvement in the audit design, we began by creating a set of simple data visualisations to stimulate discussion between ourselves, drivers and organisers. We used the DSAR Guidance document provided by Uber to identify key tables and fields, cross-referencing against other sources including weekly driver earnings reports, and sense-checking our interpretation of the fields with organisers and drivers. This side-by-side exploration allowed us to better understand the data schemas, as well as map the sensitivity and contestation around particular data types. We then identified some key metrics of interest to see if they could be calculated from the DSAR data. These included monthly net income, pay per hour (distinguishing between Uber's definition of working time, and the contrary one of the Employment Tribunal); utilisation rates (how much time is spent en route to pickup, on trip, or on standby waiting for a dispatch); and the `take rate' (how the passenger's fare was split between Uber and the driver).

This process culminated in a session with 15 drivers based in the UK, who are members of WIE and had expressed interest in learning from their work data. The session was co-led by one researcher and one organiser, who briefly explained the data, outlined potential directions for the audit, and solicited feedback and discussion. Following \cite{gallagher2023digital}, the purpose of this session was to put into practice `participatory worker data science', by helping us to identify competing interpretations of the data and to explore alternative metrics. This session helped us to refine the key areas of focus for the audit, which are presented below. We include quotes from drivers to illustrate the motivation behind each focus area:



\begin{itemize}
    \item \textbf{Pay Trends} Participants showed particular interest in graphs showing trends in monthly net pay. Reacting to stagnant or downward trends, participants said these could help them decide whether to keep working for Uber, or give them of forewarning of when they might be forced to find alternative work.
    \item \textbf{Working time definition} Despite some initial confusion, most participants understood and valued a graph displaying pay per hour in two ways: first, using Uber's definition of working time, and second, using the alternative expanded working time definition from the Employment Tribunal. As one driver observed of Uber's definition, \emph{`it's only from the time you accept the job until you drop off, that's for them what's considered working time period'}, whereas the line on the graph representing the Tribunal's working time definition showed the `real' amount of work, and was referred to as \emph{`the real line'}. We therefore sought to measure the difference between pay per hour on both definitions in our analysis.
    \item \textbf{Acceptance rates} Participants raised several points in relation to the acceptance rate data (i.e. what proportion of jobs offered were accepted by the driver). One hypothesis was that there would soon be a division between more experienced, discerning drivers, for whom acceptance rates may be lower, and newer drivers, who would accept poorly paid jobs. They argued that the pay is now worse because of a recruitment drive creating excess supply. This highlighted the collective impacts of different acceptance/rejection strategies. The extent to which acceptance was really voluntary was brought into question given suspicions about Uber's dispatch algorithm treating them differently based on past acceptance rates: \emph{`It's indirectly forcing the driver to accept; the natural instinct is `oh I'm not getting any dispatches, is it because I just rejected two just now? Is the system trying to ignore me? Oh I'd better accept whether it suits me or not'}. This motivated us to study any relationships between pay per hour and acceptance rates.
    \item \textbf{Commission / Take Rate} Many drivers lamented the loss of key information that has previously (before dynamic pricing) been provided in their weekly earnings report: \emph{`One thing they were showing before was the price the client was paying, and now they don't. That's why now I question how much they charge the client and how much they give to us, because we cannot see}'. While Uber claimed to only take 25\%, many were sceptical. One stated: \emph{`they are actually charging more than 25\% ... They want to hide how much commission they are taking from you'}. Some described circumventing this by asking customers how much they had paid: \emph{`The lady paid £64 but I was getting exactly 50\% which was £32}'. Such practices were seen as socially awkward (and have since been banned by Uber), but the only way they could see to uncover potentially unfair pricing for the driver and the customer. As one participant put it: \emph{`That's when you discover they [Uber] are robbing us and the customer'}.
    \item \textbf{Utilisation}. Some drivers were concerned about the seemingly increasing amount of standby time (i.e. time on the app, ready and willing to work, but waiting for a dispatch).  As one driver pointed out, \emph{`it doesn't cost Uber anything to have every one of us on the road at the same time, regardless how much work there is}'. Many agreed that the recent influx of new drivers - \emph{`10,000 new drivers on the road'}, as one highlighted - was leading to more competition for jobs and therefore more standby time. One participant argued that more conscious utilisation of the workforce would benefit drivers, suggesting that Uber \emph{`should introduce certain hours, they could advise drivers that you could do this, or this'}, lowering standby time and helping drivers plan their working hours around demand. Another driver agreed, adding that \emph{`they invest a lot in AI, but they don't invest in our tools; we are the ones who need to have that sort of information'}. This motivated the inclusion of utilisation rate analysis in our audit.
    \item \textbf{Dynamic pricing} The topic of dynamic pricing itself was brought up at multiple points during the session, which motivated our choice to analyse impacts before and after its introduction. The workshop took place a few months after the introduction of dynamic pricing. Drivers had many concerns about how this would change their jobs, and hypotheses about its impacts. Drivers complained about their lack of a say about such changes: \emph{`we all agreed [to dynamic pricing], to many things, because we do not have any other option'}.
\end{itemize}

Based on these identified focus areas, we continued working with WIE to narrow down the design of our audit. Key metrics included pay, take rate, acceptance rates, and utilisation. In particular, for all of these factors, we were interested in any changes before and after the introduction of dynamic pricing. This was first announced in London (where the majority of drivers in our dataset work) in February 2023\cite{uber_introduces_dynamic}. While there has been no official communication from Uber about the exact timing of rollout in different regions, based on driver testimony and information from WIE, rollout was completed across the UK by summer 2023.

For the audit, we restricted our selection to DSAR responses received between October 2023 and December 2024. This includes over 1.5 million trips in total, and at least 100k trips for each year between 2016-2024.\footnote{Code and data for this analysis are available at \url{github.com/OxfordHCC/FAccT_25_Not_Even}} We collected driver demographics including gender and age. Participants were split 96\% men, 4\% women. In terms of age, most were below 40 (\textit{20-29}: \textbf{23\%}, \textit{30-39}: \textbf{35\%}, \textit{40-49}: \textbf{28\%}, \textit{50+}: \textbf{14\%}). We compare these demographics to alternative sources to assess how representative our sample was to the whole population. While few sources show demographics of Uber drivers specifically, the regulator Transport For London \cite{tfl2023demog} report demographics for the private vehicle hire sector as a whole (which includes Uber and other private hire vehicles, but excludes traditional black cab taxis). TfL figures show a slightly higher proportion of men (97\%), and skew older (\textit{20-29}: \textbf{6\%}, \textit{30-39}: \textbf{26\%}, \textit{40-49}: \textbf{36\%}, \textit{50+}: \textbf{32\%}). A 2019 study, co-authored by an Uber employee with access to randomly sampled anonymised data from Uber's administrative records, reports 99\% drivers are men, and 52\% aged under 40 (compared to 58\% of our sample). Our sample appears to skew slightly younger than these sources, and is similar to both in terms of gender split.

\section{Audit Findings}

\subsection{Pay}
\label{subsection:pay}

\begin{figure}
    \includegraphics[width=1\linewidth]{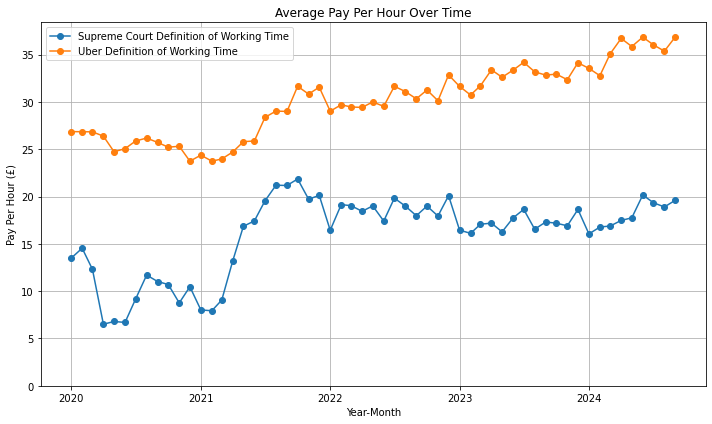}
    \caption{Pay per hour according to employment tribunal definition of working time (blue) vs Uber's definition (orange)}
    \label{fig:pay_per_hour_utilisation}
    \Description{Line chart titled displaying two data series from early 2020 to mid-2024. The y-axis shows "Pay Per Hour (£)" ranging from 5 to 40, and the x-axis shows "Year-Month". A blue line with circle markers represents pay calculated using the "Employment Tribunal Definition of Working Time". It starts around £13, dips to about £6 in mid-2020, rises to over £20 in mid-2021, and fluctuates between £16 and £20 through 2024. An orange line with circle markers represents pay using the "Uber Definition of Working Time". It starts around £27, dips slightly in early 2021, then steadily increases, peaking above £36 in 2024. A legend in the top left distinguishes between the two definitions of working time.}
\end{figure}

\begin{figure}
    \includegraphics[width=1\linewidth]{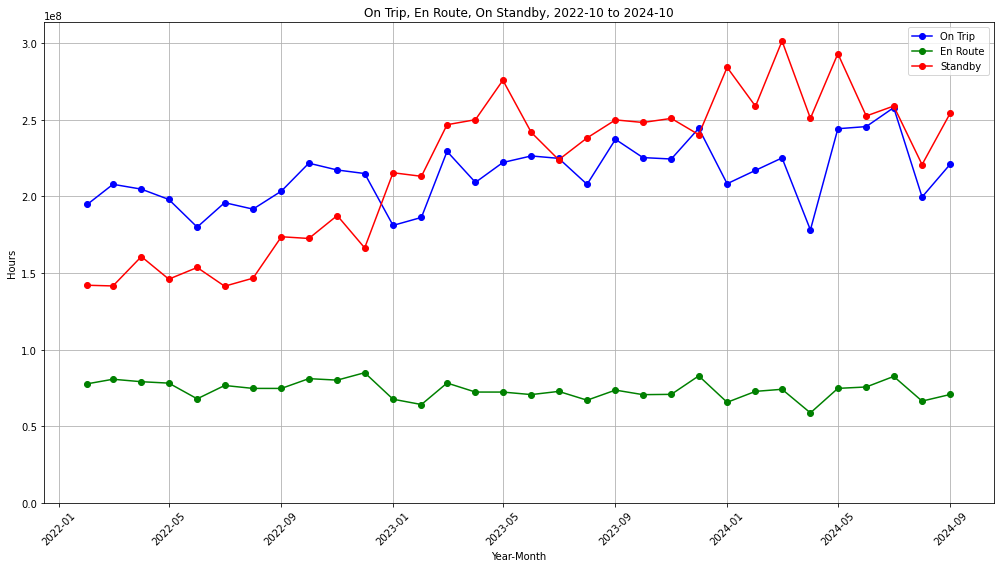}
    \caption{Average hours/day on standby (red), en route (green), on trip (blue)}
    \label{fig:standbytime}
    \Description{Line chart titled showing the total hours for three activity types over time. The x-axis represents time by "Year-Month" from early 2022 to late 2024. The y-axis shows hours ranging from .5 to 3. A blue line (On Trip) starts around 2 hours, fluctuates slightly with peaks near 2.3 hours, and dips around mid-2024. A green line (En Route) consistently stays between .3 and .6 hours with minor fluctuations. A red line (Standby) begins just under 1.5  hours, gradually increases over time, surpassing On Trip hours by early 2023, and peaks at around 3  hours several times through 2024. A legend in the top right identifies each line by activity type.}
\end{figure}

The DSAR responses contain a primary \texttt{Payments.csv} file including a wide variety of incoming and outgoing payments associated with trips. To calculate a per hour pay figure for a given week, we aggregated all incoming and outgoing payments and divided that by hours worked in the week. This is imperfect, because some kinds of payments (and driver charges) may cover periods longer than a week, or occur very irregularly. However, it is assumed these will even out when aggregated over time and across drivers.

We calculate hours worked using two definitions. The first is that favored by the Employment Tribunal: any time spent on the app ready and available to work, whether that is waiting for a request (`standby'), en route to pick up, or on trip. The second is Uber’s preferred definition: only time spent en route or on trip. We found a large difference between per hour pay depending on the definition of working time (see figure \ref{fig:pay_per_hour_utilisation}). On Uber's definition, the average is £29.46 across 2020-2024; whereas on the Employment Tribunal's definition, pay per hour was only £15.98. Both averages do not account for the costs incurred by drivers like vehicle maintenance, insurance, and fuel. 

After adjusting for inflation, average pay per hour has also declined according to both measures of working time. On the Employment Tribunal definition, average pay per hour the year prior to dynamic pricing was £22.20 (adjusted for inflation), vs an average of £19.06 the year post-dynamic pricing. On Uber's definition, average pay per hour the year prior to dynamic pricing was £37.01 (adjusted for inflation), vs an average of £35.91 the year post-dynamic pricing.\footnote{Inflation figures come from the UK Office for National Statistics document `RPI All Items: Percentage change over 12 months'} We also note that average pay per hour on Uber's working time definition is less turbulent around year 2020, where demand may have been lower due to the COVID pandemic public health measures which restricted social mixing. At such times, drivers bear the cost of low demand, as they drive around waiting for trips. While this shows up in the low pay per hour according to the Employment Tribunal definition, it is less visible under Uber's definition, where working time only accounts for time en route to pickup or on trip.

Note that these figures are averaged across all drivers and trips, and therefore may not reflect the experience of individual drivers. As shown below in section \ref{section:inequality}, pay per hour has declined more significantly for drivers who have been driving consistently the years before and after the introduction of dynamic pricing; the averages reported here for the year after dynamic pricing are higher due to the higher rates experienced by drivers who joined after dynamic pricing.

\subsection{Utilisation Rates}

Figure  \ref{fig:standbytime} shows the average hours per day spent on trip, en route, and on standby, aggregated across all drivers in a given month. While on trip and en route time has remained stable, standby time has risen significantly - increasing by over 1 hour a week since 2022, and remains high. In most months since 2023, drivers spend more time waiting to be allocated their next job than they do on journeys with passengers.

\subsection{Commission / Take Rates / Surplus}

\begin{figure}
    \includegraphics[width=1\linewidth]{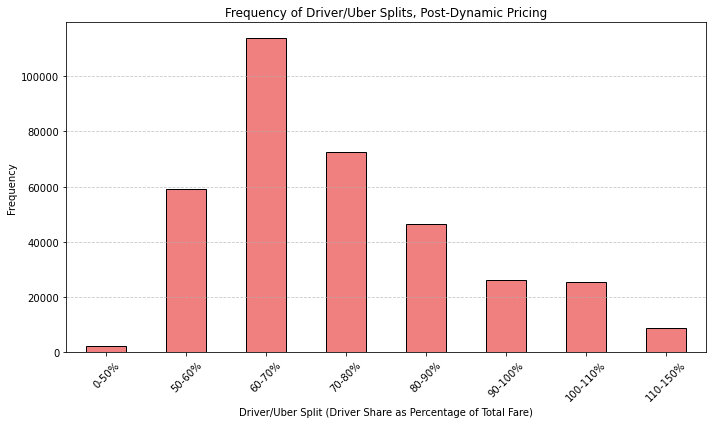}
    \caption{Frequency of driver/Uber splits, post dynamic pricing}
    \Description{A bar chart showing the frequency distribution of driver/Uber fare splits after dynamic pricing implementation. The x-axis shows driver share as percentage ranges of total fare (0-50\%, 50-60\%, 60-70\%, 70-80\%, 80-90\%, 90-100\%, 100-110\%, 110-150\%), while the y-axis shows frequency from 0 to 120,000. The distribution is right-skewed with the highest frequency of approximately 112,000 occurrences in the 60-70\% range, followed by about 73,000 in the 70-80\% range. Lower frequencies are observed at the extremes, with very few instances below 50\% or above 110\% driver share}
    \label{fig:frequency_take_rate}
\end{figure}

Until the introduction of dynamic pricing, Uber had publicly advertised their commission rates at 25\% (having risen from 20\%)\cite{bbc_impossible}. However, since dynamic pricing, the customer fares were removed from drivers' weekly earnings report, meaning drivers could no longer see how much of a cut Uber takes from any given trip (the `take rate'). As explained above, drivers are banned from asking passengers about their fares, making it impossible for them to work out the take rate. However, customer fares were included in DSAR responses, albeit in an incomplete form. A unique trip identifier is not provided in the data, but by joining multiple tables and inferring which payment related to which trip based on timestamps, we were able calculate the take rate by dividing the `original fare' by the driver payment associated with the trip.

However, this integration was not possible for one year's worth of data due to an apparent change in the backend system which generates the DSAR data. Prior to February 2022, DSARs contained an `original fare' field in \texttt{Trips.CSV}, which corresponded to the price paid by the customer (we independently verified this with individual drivers); as well as a `commission' charge, which appeared at the same time as a corresponding trip, in \texttt{Payments.CSV}, which always added up to the correct commission rate (20\% prior to 2021, and 25\% after). However, in February 2022, the `original fare' appeared to no longer represent the customer's fare, and the commission charges disappeared from \texttt{Payments.CSV}. This coincides with the point at which Uber introduced measures to (partially) comply with the previous year's UK Supreme Court ruling. Explaining these compliance measures, Uber stated: \emph{`the rider }[i.e. the passenger] \emph{will now pay Uber directly instead of paying the driver - a service fee is no longer suitable as a mechanism to provide payment to Uber for using our app'}.\footnote{\url{https://www.uber.com/en-GB/blog/driver-terms-faq/}}. This means that take rate data is missing from our dataset from 2022-02. However, since the introduction of dynamic pricing in 2023-02, the original fare field appears to have been restored to its previous source, once again reflecting the fare paid by the customer (as confirmed by drivers and customers), enabling us to continue our take rate analysis. As discussed above, in January 2025 Uber begun to reveal weekly average take rates in the app, but still do not disclose trip-level take rates.



Figure \ref{fig:frequency_take_rate} shows the distribution of take rates (in bins of 10\%) for all trips since the introduction of dynamic pricing. In contrast to the standard 75/25\% split prior to dynamic pricing, driver take rates are now frequently as low as 50-60\%, and in some cases even below 50\%. Conversely, in other cases, the driver's take rate can exceed 100\%, i.e. the passenger's fare does not even cover the driver's pay, meaning Uber makes a loss on these journeys.

Mean average take rates have remained at 75\%, which supports Uber's claim in \cite{bbc_impossible} that their 25\% cut remains stable. However, the distribution is skewed right, with the median average take rate post-dynamic pricing at only 71\%. Only 46\% of drivers have managed to maintain an average take rate of 75\% or above.

Note that Uber calculates its take rate differently in the weekly averages provided to drivers in the app, because it subtracts passenger promotions and third party fees from its take rate. We did not include these in our take rate analysis for several reasons. First, passenger promotions are discounts that Uber has decided to give to passengers; as such, the full price should be considered as Uber's `real' valuation of the trip, not the discounted amount paid by the passenger. Second, third party fees are not driver pay, but reimbursements of operating costs (such as London's congestion zone charge) that drivers have had to pay out of their own pocket (which also aren't included in our analysis). Finally, we were also unable to link these costs to particular trips because Uber does not provide a trip ID, and unlike with direct trip payments, we could not infer links by correlation of timestamps.

\begin{figure}
    \includegraphics[width=1\linewidth]{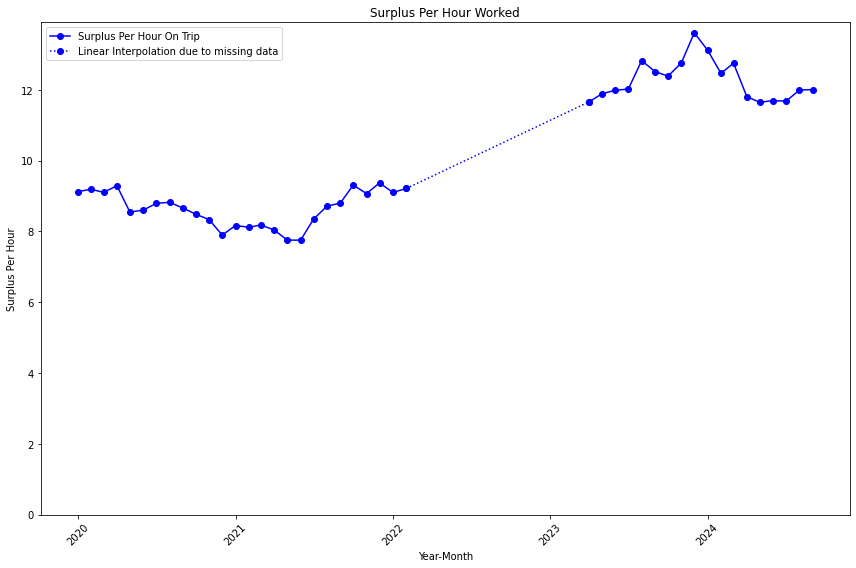}
    \caption{Surplus generated by drivers for Uber per hour on trip}
    \Description{A line chart showing surplus per hour worked from 2020 to 2024. The y-axis shows surplus per hour ranging from approximately 7.5 to 13.5, while the x-axis shows time periods by year-month. The chart displays two types of data points: solid blue circles connected by lines representing actual 'Surplus Per Hour On Trip' data, and dotted blue circles representing 'Linear Interpolation due to missing data.' The trend shows relatively stable values around 8-9 from 2020 through early 2022, followed by a significant upward trend starting in late 2022, reaching peak values around 13.5 in late 2023, then declining and stabilizing around 12 in 2024. There's a notable gap in the data during 2022 where linear interpolation was used to estimate missing values.}
    \label{fig:surplus}
\end{figure}

In addition to measuring Uber's take rate, we also calculated the average surplus, defined as the amount of revenue generated for Uber by each driver for each hour they are on a trip, minus the driver's labour cost (see Figure \ref{fig:surplus}). This is not equivalent to surplus profit, as it does not account for Uber's non-driver costs (such as employees, infrastructure, overheads, etc). However, assuming such costs are relatively low per trip, and have remained stable, this provides a reasonable proxy of the change in the surplus value generated by a driver while en route / on trip. We find that Uber's average surplus per worker hour (minus non-driver costs) has increased by 38\%, from an average of £8.47 in the period 2020/01-2022/02, to an average of £11.70 in 2023/03-2024/10.

\begin{figure}
    \centering
    \includegraphics[width=1\linewidth]{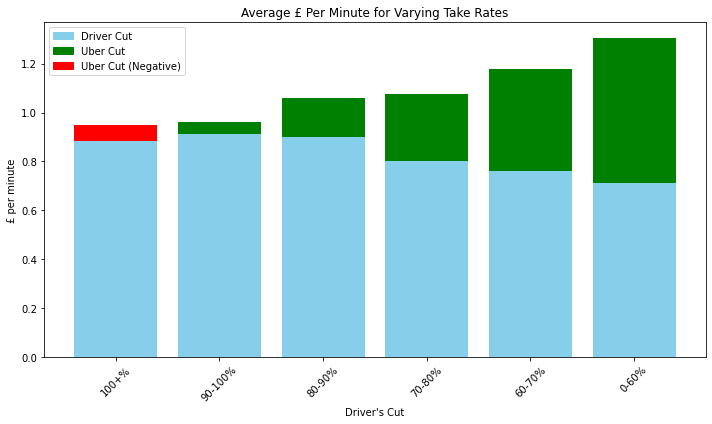}
    \caption{Average per minute fare for varying driver/uber splits. As Uber's cut increases, fares increase, but drivers earn less per minute (on trip) in absolute terms. Red area shows Uber's cut as negative, i.e. when driver's pay is more than 100\% of the fare}
    \label{fig:driver_uber_split}
    \Description{A stacked bar chart showing average earnings per minute (in British pounds) for different driver cut percentages. The x-axis shows driver's cut ranges from 100\% down to 0-60\%, while the y-axis shows earnings per minute from 0 to about 1.4 pounds. Each bar is divided into three components: light blue on the bottom representing 'Driver Cut', dark green in the middle representing 'Uber Cut', and red on top representing 'Uber Cut (Negative)'. It is only negative in the first "100\%+" bar." The total bar height increases as the driver's cut percentage decreases, ranging from about 0.95 pounds per minute at 100\% driver cut to about 1.3 pounds per minute at 0-60\% driver cut. The driver portion (light blue) decreases as driver cut percentage decreases, while Uber's portion (green) increases correspondingly.}
\end{figure}

Given that mean average take rates have remained stable, and median take rates have only decreased by 4 percentage points, this increase in surplus cannot be explained by increasing average take rates. However, looking at the distribution of take rates across different value trips provides another explanation (see Figure \ref{fig:driver_uber_split}). We find that Uber's take rates are not evenly distributed across customer fares; rather, the higher the fare charged to the customer, the higher Uber's take rate, and the less drivers earn per minute in absolute terms. Drivers are therefore worse off when they select higher-value fares.

\subsection{Inequality Between Drivers}
\label{section:inequality}

\begin{figure}
    \includegraphics[width=1\linewidth]{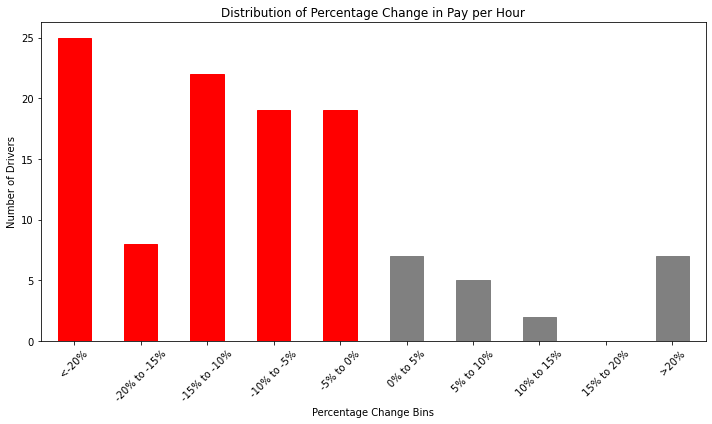}
    \caption{Distribution of percentage change in pay per hour after dynamic pricing, for higher earners (grey) and lower earners (red)}
    \Description{A histogram showing the distribution of percentage change in pay per hour among drivers. The x-axis shows percentage change bins ranging from less than -20\% to greater than 20\%, while the y-axis shows the number of drivers from 0 to 25. The chart uses red bars for negative changes and gray bars for positive changes. The distribution is heavily skewed toward pay decreases, with the highest frequency of about 25 drivers experiencing decreases of less than -20\%. Other notable frequencies include approximately 22 drivers with -15\% to -10\% decreases, 19 drivers each with -10\% to -5\% and -5\% to 0\% changes, and 8 drivers with -20\% to -15\% decreases. Positive changes are much less common, with about 7 drivers experiencing 0\% to 5\% increases, 5 drivers with 5\% to 10\% increases, 2 drivers with 10\% to 15\% increases, and 7 drivers with increases greater than 20\%. No drivers experienced 15\% to 20\% increases.}
    \label{fig:distribution_percentage_change}
\end{figure}

\begin{figure}
    \includegraphics[width=1\linewidth]{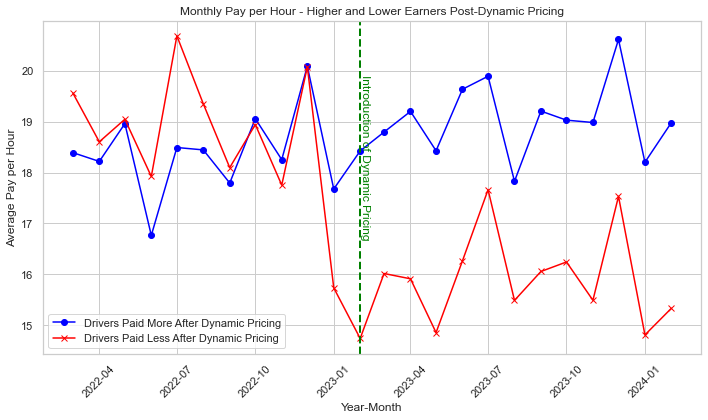}
    \caption{Pay per hour for higher (blue)/lower (red) earners after dynamic pricing.}
    \Description{A Line chart showing monthly pay per hour for two groups of drivers from 2022-05 to 2024-09, with a vertical green dashed line marking 'Introduction of Dynamic Pricing' around 2023-01. The chart compares 'Drivers Paid Same or More After Dynamic Pricing' (blue line with circles) and 'Drivers Paid Less After Dynamic Pricing' (red line with X markers). The y-axis shows average pay per hour ranging from about 14.5 to 22 pounds. Before dynamic pricing, both groups showed similar pay patterns around 18-20 pounds per hour with some fluctuation. After dynamic pricing implementation, the two groups diverge significantly: drivers paid the same or more maintain higher earnings (generally 18-22 pounds per hour with a peak near 22), while drivers paid less show consistently lower earnings (generally 15-19 pounds per hour). The gap between the two groups widens over time, with the higher-earning group reaching peak values around 2024-04 before declining slightly.}
    \label{fig:inequality}
\end{figure}

How are changes in pay post-dynamic pricing distributed between drivers? We selected the period covering one year prior to and one year after the introduction of dynamic pricing (2022/02-2024/02), and selected data from only those drivers who had monthly data covering this period (114 drivers).

We found that average per hour pay for this subset of 114 drivers had decreased from £18.52 to £17.07 (a larger drop than that observed across all drivers in the dataset above). We found that 93 drivers were worse off in terms of their average per hour pay after dynamic pricing, and 21 were better off (Figure \ref{fig:distribution_percentage_change}, left). Figure \ref{fig:inequality} (right) shows rolling monthly per hour pay averages for those drivers whose average per hour pay was lower (red), and higher (blue). This shows that these two groups had relatively similar pay per hour pre-dynamic pricing, but after its introduction in 2023/02, they consistently diverge (with some re-convergence towards the end of 2024). This mirrors a similar pattern found in \cite{calacci2022bargaining}, where the introduction of a new algorithm on the Shipt gig work platform lead to inconsistent effects on pay among gig workers.

\begin{figure}
    \includegraphics[width=1\linewidth]{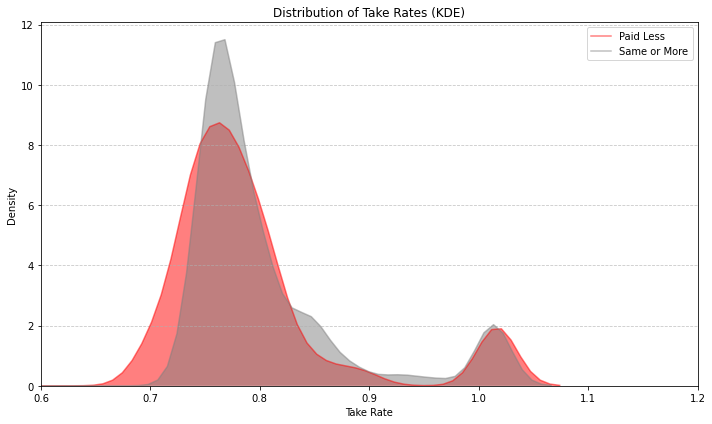}
    \caption{Take Rates for Lower-Paid vs Same/Higher-Paid}
    \label{fig:take_comparison}
    \Description{A Kernel Density Estimate (KDE) chart titled "Distribution of Take Rates (KDE)" showing the distribution of Uber's take rate for two driver groups after dynamic pricing was implemented: those who were Paid Less (in red) and those who were Paid the Same or More (in gray). The x-axis represents the Take Rate, ranging approximately from 0.6 to 1.2. The y-axis shows Density, peaking around 11. The chart shows that both groups have a major peak between 0.75 and 0.8, indicating this was the most common take rate. However, the Paid Less group (red) has a slightly broader distribution with a secondary smaller peak around 1.02, suggesting a subset of drivers faced higher take rates. The Same or More group (gray) has a sharper peak. A legend in the upper right identifies the two groups by color.}
\end{figure}

\begin{figure}
    \includegraphics[width=1\linewidth]{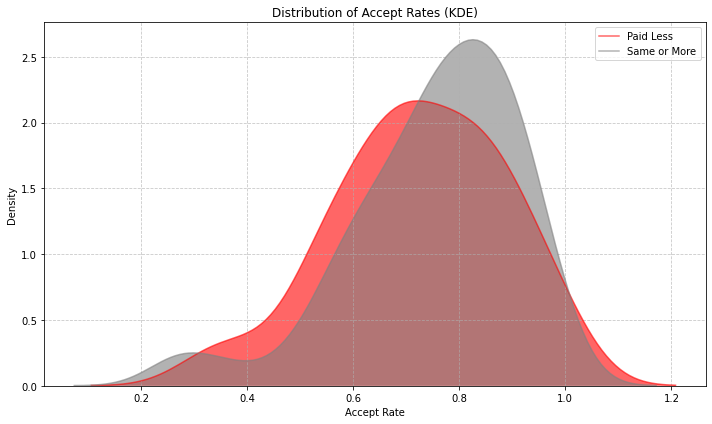}
    \caption{Accept Rates for Lower-Paid vs Same/Higher-Paid}
    \label{fig:accept_comparison}
    \Description{A Kernel Density Estimate (KDE) chart titled "Distribution of Accept Rates (KDE)" showing the distribution of ride accept rates for two groups of Uber drivers after dynamic pricing was implemented: drivers who were Paid Less (red) and those who were Paid the Same or More (gray). The x-axis shows Accept Rate, ranging from 0 to 1.2. The y-axis shows Density, with a peak just over 2.5. The Paid Less group (red) has a broader distribution, peaking around a 0.7 accept rate, with more density toward lower accept rates (below 0.6). The Same or More group (gray) is slightly more concentrated, peaking near 0.85, indicating they tended to accept slightly more ride offers. A legend in the upper right corner labels the two groups.}
\end{figure}

Based on discussion of this data with drivers and WIE, we undertook some further exploratory analysis to understand if there are any factors that might explain the differences between these groups (those who were earning less, vs same or more). We found no clear differences in hours worked, or number of years experience (see Appendix, Figure \ref{fig:lower_comparison_no_diff}); however, we saw some differences in driver take rates and acceptance rates. For those who earned the same or more per hour, there is a higher concentration of trips with a driver take rate around 76\%, while for those paid less, the distribution is wider (Figure \ref{fig:take_comparison}). In terms of accept rates, those paid the same or more skew further right (Figure \ref{fig:accept_comparison}). While neither set of differences are large, when combined they may at least partially explain the differences in pay per hour.

\subsection{Predictability of Pay}\label{predictability}
A driver's livelihood depends on their ability to guess what kinds of trips they will get at particular times and places, and how much those trips will pay. However, in our participatory worker data science sessions, drivers frequently complained about the unpredictability of pay post-dynamic pricing. We sought to measure this unpredictability by training and testing models to predict the pay associated with a trip, based on over 60 variables associated with each trip. These included duration, distance, time of day, day of week, airport trips, etc. The purpose of this exercise was to test whether, even with a large amount of pooled data, the distribution of pay under dynamic pricing would become unpredictable based on pre-dynamic pay patterns. Our model performance represents an upper bound on the mental models and tacit knowledge built up by drivers over the years, to guide their strategies about when and where to work. A reduction in performance of this model would imply a reduction in the utility of such knowledge under the new dynamic pricing regime.

First, we trained linear regression models on subsets of subsequent years, and then tested them on the following year.
We evaluate the regression models on a test set with R-squared (out of 1.0).
In Table \ref{tab:pred_one}, we used the previous year ($Y-n$) to predict the current year $Y$ (e.g. train on 2021, test on 2022).
This would evaluate how consistent the variables from $n$ years ago would be on a given year. We see a drop in performance in 
2020 (after the Covid-19 pandemic), and an even more significant drop post-2023 (after the introduction of dynamic pricing). Performance in previous years improved or stayed steady year on year, up until these years where it dropped off. 
For example, if we used any data prior to 2022 to predict pricing on 2023-4, the r2 score remains consistently poor/negative.
In Table \ref{tab:pred_all} (see Appendix), we used all data up until the given year $1, ..., Y-n$ (e.g. train on data from 2014-2021, test on 2022).
We find a similar pattern as above, indicating even consistent variables over multiple years do not predict well post-dynamic-pricing.

We note that models that are both trained and tested on post-dynamic pricing do perform well. So it is possible that over time, if the dynamic pricing model and other factors do not change significantly, drivers may be able to gradually rebuild the tacit knowledge built up in previous years to match the new dynamic pricing regime. Nevertheless, in the interim, these changes in distribution of the underlying variables pre- and post-dynamic pricing would likely be highly disruptive to any driver whose strategies and working patterns are based on previous knowledge.

\begin{table*}[]
\begin{tabular}{c|ccccccccccc}
\multicolumn{1}{l}{} & Y     & Y-1                  & Y-2                  & Y-3                  & Y-4                  & Y-5                  & \multicolumn{1}{c}{Y-6}    & \multicolumn{1}{c}{Y-7}    & \multicolumn{1}{c}{Y-8}    & \multicolumn{1}{c}{Y-9}    & \multicolumn{1}{c}{Y-10}   \\ \hline
2014                 & 0.873 & \multicolumn{1}{l}{} & \multicolumn{1}{l}{} & \multicolumn{1}{l}{} & \multicolumn{1}{l}{} & \multicolumn{1}{l}{} &                            &                            &                            &                            &                            \\
2015                 & 0.628 & 0.485                & \multicolumn{1}{l}{} & \multicolumn{1}{l}{} & \multicolumn{1}{l}{} & \multicolumn{1}{l}{} &                            &                            &                            &                            &                            \\
2016                 & 0.799 & 0.592                & 0.211                & \multicolumn{1}{l}{} & \multicolumn{1}{l}{} & \multicolumn{1}{l}{} &                            &                            &                            &                            &                            \\
2017                 & 0.849 & 0.817                & 0.134                & -0.410               & \multicolumn{1}{l}{} & \multicolumn{1}{l}{} &                            &                            &                            &                            &                            \\
2018                 & 0.887 & 0.881                & 0.858                & -0.171               & -0.615               & \multicolumn{1}{l}{} &                            &                            &                            &                            &                            \\
2019                 & 0.871 & 0.865                & 0.856                & 0.803                & -1.583               & -2.008               &                            &                            &                            &                            &                            \\
2020                 & 0.919 & -0.053               & 0.781                & 0.874                & 0.491                & -7.685               & \multicolumn{1}{c}{-7.766} &                            &                            &                            &                            \\
2021                 & 0.821 & 0.801                & 0.283                & 0.509                & 0.772                & 0.517                & \multicolumn{1}{c}{-7.808} & \multicolumn{1}{c}{-6.277} &                            &                            &                            \\
2022                 & {\color{red}0.861} & 0.837                & 0.808                & 0.397                & 0.731                & 0.781                & \multicolumn{1}{c}{0.627}  & \multicolumn{1}{c}{-4.477} & \multicolumn{1}{c}{-3.678} &                            &                            \\
2023                 & 0.858 & {\color{red}-54.023}              & -20.144              & -5.625               & -0.739               & -9.888               & \multicolumn{1}{c}{0.787}  & \multicolumn{1}{c}{0.679}  & \multicolumn{1}{c}{-3.219} & \multicolumn{1}{c}{-3.441} &                            \\
2024                 & 0.891 & 0.886                & {\color{red}-0.219}               & 0.434                & 0.711                & 0.833                & \multicolumn{1}{c}{0.630}  & \multicolumn{1}{c}{0.828}  & \multicolumn{1}{c}{0.774}  & \multicolumn{1}{c}{-3.489} & \multicolumn{1}{c}{-3.674}
\end{tabular}
\caption{Predicting price at year $Y$ using previous year $Y-n$}
\label{tab:pred_one}
\Description{A table showing correlation coefficients for predicting Uber ride prices in year Y using data from previous years (Y-1 through Y-10) from 2014 to 2024. The table has years 2014-2024 in rows and lag periods Y through Y-10 in columns. Values represent predictive correlations, with higher positive values indicating better predictability. Key findings highlighted in red show that 2023 data becomes highly unpredictable using 2022 data (correlation of -54.023), and 2024 prices show poor predictability using 2022 data (correlation of -0.219). Prior to this period, correlations were generally positive and moderate (typically 0.4-0.9), indicating stable predictability. The dramatic shift to negative correlations around 2022-2023 suggests that the introduction of dynamic pricing significantly disrupted the historical price patterns, making previous years' data less reliable for price prediction.}
\end{table*}

\section{Discussion and Conclusion}

Our discussion is divided into direct commentary on the audit findings and limitations of our approach, followed by various implications for political action, policy, and research.

\subsection{Audit Findings}
Our findings suggest that post-dynamic pricing, many aspects of Uber drivers' jobs have gotten worse. Average pay per hour on the app is stagnant, and is lower in real terms in the year following the introduction of dynamic pricing. The discrepancy in pay per hour between the Employment Tribunal's definition of working time vs Uber's has increased. Uber's median take rate per driver has increased from 25\% to 29\%, and on some trips the take rate is over 50\%. Furthermore, the higher take rates are concentrated among higher-fare trips, which explains how Uber can extract an additional 38\% surplus from its driver's labour on average (equivalent to £3.23 per hour), a phenomenon which is obscured in the average take rate. Many drivers are earning substantially less per hour. Utilisation has also changed; in some months, drivers spend more time waiting for the next trip than they do actually on trip. Finally, any tacit knowledge drivers have built up over years about how much pay a given trip is likely to yield may no longer help them; as our modelling suggests, the predictability of pay drastically changed after dynamic pricing was introduced.




\subsection{Limitations}
Our study has various limitations. First, our sample of 258 drivers, while relatively large for a research study, is still small relative to the whole population of 100,000 Uber drivers in the UK. It is also drawn from drivers who are sufficiently motivated to sign up to WIE's DSAR scheme, and who may therefore be more likely to already have had problems with Uber's platform than other drivers (although, as explained above, the demographics of our sample appear similar to the wider population). While some of our observations are directly attributable to dynamic pricing (e.g. those related to the variation in take rates), our study design does not enable us to isolate the causal effect of dynamic pricing on pay, utilisation and predictability. Contemporaneous changes in passenger demand, traffic flows, workforce size and composition, and even potential behavioural change among drivers in response to dynamic pricing, limit our capacity for causal attribution.


\subsection{The Ethics of Dynamic Pay and Pricing}
In its initial incarnation, Uber presented a simple payment model to drivers. Pay was based on a flat rate per mile and per minute, and Uber took a fixed commission. Since then, the payment model became increasingly complex. Additional factors were introduced into the pricing calculation; upfront fares based on Uber's prediction of duration and distance were introduced; the commission rate increased, and then was removed; other kinds of payments and charges were introduced following the Supreme Court ruling. Finally, the entire system was replaced by dynamic pricing. As a result, the current payment model is unrecognisable from the original one. Drivers who signed up in earlier waves of recruitment, even if they didn't know exactly what jobs they'd get, could at least know how pay is calculated (a simple function of time and distance), giving them some predictability, stability and autonomy. But this is made impossible by dynamic pricing (as our prediction models attest). As Uber's CEO described the trajectory in a call to investors: \emph{`you've gone from just flat time and distance to now kind of point estimates for every single trip based on the driver... targeting of different trips to different drivers based on their preferences or based on behavioral patterns that they're showing us, that really is the focus going forward, offering the right trip at the right price to the right driver'}.\footnote{\url{https://www.fool.com/earnings/call-transcripts/2024/02/07/uber-technologies-uber-q4-2023-earnings-call-trans/}}

A key feature of the old pricing model was that it forced Uber to be transparent about its cut, as it was a fixed commission across all trips. This meant if it wanted to charge a customer more, the driver would always automatically benefit proportionally, thus aligning driver's interests with those of the platform, against those of the customer. Drivers were unlikely to complain about higher customer prices if they benefit proportionally. However, post-dynamic pricing, Uber has made commission take rates variable, cutting the tie between customer price and driver pay. As our analysis of Uber's surplus shows, Uber also take a higher percentage of the most valuable fares, thus extracting even greater surplus per hour a driver works.

For drivers, this breaks what was a mutually beneficial compact that had previously kept them from making common cause with customers. As drivers in our participatory data science sessions argued, drivers and customers are now both being `robbed'. One recalled: \emph{`You ask the customer ... how much are you being charged? And they tell you this much, and then you think well I'm only getting this much ... I've challenged Uber before ... What they say to me is: you saw the upfront price, you took it, that means you accept it. They just blocked me, they shut it down.'}. Instead of the dependability of being paid for the time and gas spent on trip, minus a consistent and publicly known commission, drivers must now continuously gamble, never sure how much cut Uber is taking. This reflects a broader development that legal scholar Veena Dubal has called the `algorithmic gamblification' of work, where: \emph{`on-the-job data collection and algorithmic decision-making systems ... are undermining the possibility of economic stability and mobility through work by transforming the basic terms of how workers are paid'}\cite{dubal2023house}. In this model, the house always wins.




\subsection{The Reserve Army of Platform Labour}
Under capitalism, workers have the freedom to sell their labour at whatever price employers are willing to pay. However, the existence of a `reserve army' of unemployed labourers, who would be willing to sell their labour for cheap, is key to resisting pressure to increase wages from those who are in employment \cite{marx1867}. Since the rise of zero-hours contracts, this distinction between the reserve army and the rest of the workforce was already blurred, with a large class of variably (un)employed workers. But gig work platforms have turned this dynamic between the employed and reserve army into an increasingly real-time, variable proposition. Uber has brought many new drivers onto the platform. When they get there, they are added to the waiting list of drivers on standby, looking for work. Faced with such oversupply, drivers become reluctant to turn down offered trips, even if low paid, because they fear other drivers will accept them.

Having this excess supply adds very little additional cost to Uber, but does cost the worker; as one driver put it, \emph{`They have no cost, but we have cost'}. Such costs include: time wasted while on standby, looking for work, fuel, toll charges, as well as upfront costs of getting set up as a driver on the platform in the first place. Excess supply also has social and environmental costs; more drivers driving around looking for work means greater carbon emissions and pollution on the streets; greater congestion, slowing down other traffic and public transit; and safety impacts associated with more cars on the road. All of these costs are fully externalised by Uber, providing no incentive for them to fix them.

As drivers in our participatory data science sessions recognised, the ease with which customer requests are fulfilled might appear to be down to Uber's sophisticated AI, but may actually have more to do with the oversupply of labour: \emph{`they're trying to create this perception so the customers think `Oh yeah, it's good that they've used [AI], because we couldn't get a car before, now we get a car just like that ... but there's a lot more drivers out there just sitting around!'}. As drivers pointed out, there could be other, more equitable ways to use AI to better allocate the useful work of transporting passengers around, that would provide stable and dependable work. As shown in section \ref{predictability}, demand for transportation is to a large degree predictable, even if dynamic pricing isn't. A transportation service with access to such data at scale could enable better prediction and co-ordination, allowing drivers to work at pre-arranged set hours, greatly reducing unnecessary standby time and the social and environmental costs it brings to drivers, cities, and the environment \cite{bates2018transforming}.

\subsection{DSARs for algorithm auditing and worker data science}
To our knowledge, this is the first large-scale algorithm audit performed using DSARs, other than the aforementioned examples of the OpenSCHUFA Credit Score initiative. One of our goals was to investigate whether, despite previously identified limitations, DSARs have the potential to be effective as tools for algorithm auditing and worker organising. Our experience suggests that this can be a promising tool for research. It enables bottom-up investigations and provides ground-truth data directly from companies rather than filtered through third party APIs (e.g. \url{Argyle.com}).

However, without the sustained and long term efforts of WIE to mobilise their members to engage with DSARs, this would not have been possible. The difficulty of pressuring and negotiating with platforms themselves to respond to access requests, often requiring complaints to regulators and legal representation, cannot be underestimated. The responsibilities of handling requests, and safely stewarding the large amounts of personal data they contain, are also substantial. The use of mass DSARs as a research tool should therefore not be undertaken lightly by researchers. Researchers may need to partner with specialist organisations who can take on some of this responsibility. In addition, privacy-preserving and collectively governed research infrastructure is needed, to ensure participants can co-determine how their data is handled, and how research is turned into action \cite{boeschoten2022framework,stein2023you,zhao2023libertas}.

Given such challenges of DSAR-based audits, platform transparency should also be pursued through regulation and/or collective agreements. One example for large online platforms is the EU Digital Services Act (DSA) Article 40. More specific to rideshare platforms, related initiatives include transparency reports\cite{rao2024rideshare}, rideshare data release efforts in cities like New York and Chicago\footnote{\url{https://www.nyc.gov/site/tlc/about/tlc-trip-record-data.page}, \url{https://data.cityofchicago.org/Transportation/Transportation-Network-Providers-Drivers/j6wf-834c/about_data}}, and laws such as Colorado Senate Bill 24-75 requiring platforms to disclose fare, distance and direction information to drivers before they accept a ride. The latter was influenced by FairFare, a recent project similar to this one which measures take rates on rideshare platforms \cite{calacci2025fairfare}. Similarly, WIE have recently called on UK city governments to establish better public data on ridesharing platforms impact on labour markets, communities and environments, based in part on the analysis above \cite{wie_dying}.




\section{Conclusion}
We undertook a 2-year long participatory action research project to put the idea of worker data science into practice. Driven by the concerns, hypotheses, and strategies of workers and organisers, we undertook an audit of Uber's algorithms, finding empirical evidence for several harmful trends. The results of this audit are already being put into action by WIE, used as evidence in its organising, advocacy and policy work (see, e.g. \cite{wie_dying}). We conclude that DSAR-based auditing can play an important role in securing a better future for gig workers in the face of algorithmic injustice.


\bibliographystyle{ACM-Reference-Format}
\bibliography{references}
\newpage

\section{Endmatter}

\subsection{Author Contributions}

The project was led by Reuben Binns and Jake Stein who conducted workshops, interacted with workers one-on-one, and performed data transformation, programming, and analysis. Reuben Binns led preparation of the original draft while Jake Stein was on leave. Siddartha Datta assisted with the analysis of predictability of pay. Max Van Kleek and Nigel Shadbolt provided writing review and supervision. 


\subsection{Acknowledgements}
This work was supported by the Oxford Martin School Ethical Web and Data Architectures in the Age of AI project. Of course, it would not have been possible without the advocacy of James Farrar and Cansu Safak of Workers Info Exchange and all the workers who contributed. WIE were also supported by Open Society Foundation, Digital Freedom Fund, Trust for London, and Omidyar Network. Thanks to Abdurzak Hadi and other members of the App Drivers and Couriers Union for their collaboration with the workshops. Thanks to Anton Ekker of Ekker Law. Thanks to Jeremias Adams-Prassl, Six Silberman, and the iManage project at the University of Oxford for creating a vibrant and insightful community of scholars of algorithmic management which have greatly contributed towards our understanding of this field. Thanks to Abby Gilbert and the Institute for the Future of Work for supporting early development of the DSAR-based auditing concept. Thanks also to Vidminas Mikucionis, Hilde Weerts, Tony Curzon Price, and anonymous reviewers at ACM CHI'25 and FAccT'25 for feedback on various drafts.

\subsection{Competing Interests}
The authors report no competing interests. 

\subsection{Positionality Statement}
Our positionality to this research is layered given our interaction with Uber, workers, and WIE. As researchers, our interest in the greater patterns of algorithmic management and pay visible in aggregate data could have led to different prioritization when compared to workers' individual interests in individual data. This was balanced with the interests of collective advocacy brought by WIE, who are motivated both by advocacy for workers on an individual basis and a collective one. As researchers from the University of Oxford, we are aware of our institution's deep ties to colonialism, while many of the primary subjects (and, we hope, beneficiaries) of this research are often immigrants to the United Kingdom from formerly colonized nations. 


\subsection{Ethical Considerations Statement}
This work raised various ethical concerns. One is around the privacy of drivers whose data was included in the study. As explained in \ref{subsection:datasource}, the data is highly detailed and longitudinal, containing in some cases nearly a decade of working activities. We went to great efforts to protect such data, including: writing custom software to enable the initial data cleaning, preprocessing and extraction to take place on-premises within WIE; pseudonymising identifiers through hashing; and stripping away any unecessary fields. We also spent significant time discussing and formalising the governance around the data, including working with our institutional review board, drawing up a data transfer agreement, and working with WIE to update their data protection impact assessment and security controls prior to sharing. 

Another concern is around the potential for retaliation by the platform against individual workers or WIE. To mitigate this, we carefully discussed the remit of the study and potential scenarios. We also consulted legal experts to assess whether there would be any grounds for legal challenge, e.g. around intellectual property / trade secrets, concluding that such risks were low.

\subsection{Adverse Impact Statement}

In addition to providing scholarly insights, an important motivation for this work is to help bring about justice for workers in the gig economy. Despite this, there could be unintended / adverse impacts of our research. One might be that if advocacy based on this research succeeds in improving conditions for workers, this comes at a cost for other stakeholders. For instance, customer fares could be raised even higher, to pay for higher driver take rates. Another possibility is that, in order to reduce the problem of oversupply identified here, some workers are terminated. Finally, if forced to improve pay and conditions, the platform might decide to remove itself from the UK market, putting jobs at risk and removing a service from customers. These risks are difficult to anticipate and mitigate. However, in this work, we have attempted to show how broader efforts towards justice in the gig economy could ultimately benefit society more generally. For instance, there can be common cause between drivers and customers whose pay and prices are both being squeezed by the platform. Regarding the potential loss of work for existing drivers, we would advocate for alternative public mobility platforms, based around human needs rather than profit, which could enable the more equitable sharing of work between existing drivers. 


\newpage

\section{Appendices}

\begin{figure}
    \centering
    \includegraphics[width=0.75\linewidth]{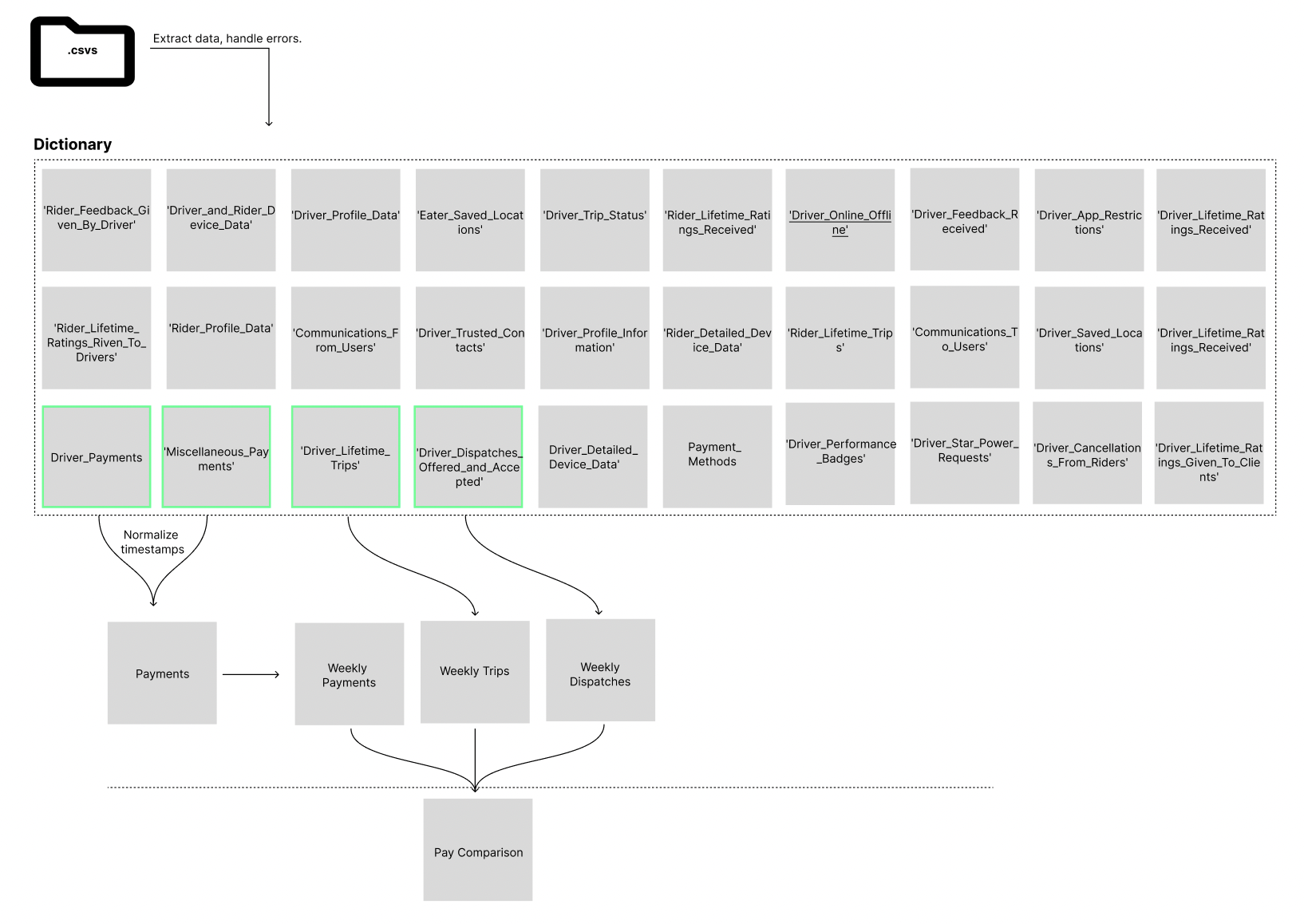}
     \Description{A table of squares surrounded by an outline labled "Dictionary" representing the data tables returned from Uber to WIE. Four squares are highlighted, "Driver_Payments", "Miscellaneous_Payments","Driver_Lifetime_Trips", and "Driver_Dispatches_Offered_and_Accepted". These squares have lines connecting them to squares labelled" Payments, Weekly Payments, Weely Trips, Weekly Dispatches and Pay Comparison."}
    \caption{Mapping data pipeline from a typical DSAR response for one aspect of our analysis (pay comparison, detailed in section \ref{subsection:pay})}
    \label{fig:DSAR}
\end{figure}

\begin{figure}
    \centering
    \includegraphics[width=0.49\linewidth]{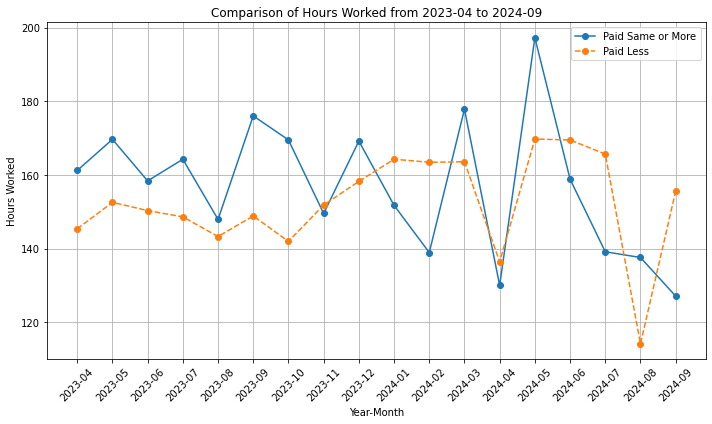}
    \includegraphics[width=0.49\linewidth]{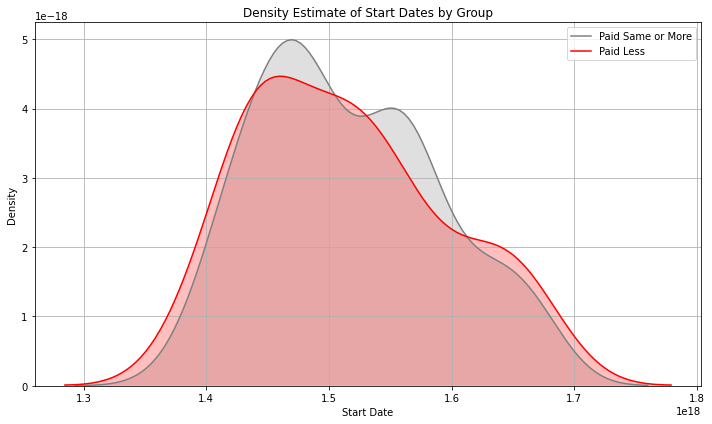}
    \caption{Left: Hours Worked, Lower-Paid (Orange) vs Same/Higher Paid (Blue); Right: Years of Experience for Lower-Paid (Red) vs Same/Higher-Paid (Grey) Drivers}
    \label{fig:lower_comparison_no_diff}
    \Description{Left: A line chart comparing hours worked from April 2023 to September 2024 between two groups of drivers: those 'Paid Same or More' (blue solid line with circles) and those 'Paid Less' (orange dashed line with circles). The y-axis shows hours worked ranging from about 115 to 200, while the x-axis shows monthly periods. Both groups show fluctuating patterns with several peaks and valleys. The 'Paid Same or More' group shows more volatile changes, with notable peaks around September 2023 (\~177 hours), February 2024 (\\~178 hours), and May 2024 (\~197 hours), and significant dips in April 2024 (\~131 hours) and September 2024 (\~127 hours). The 'Paid Less' group shows somewhat more stable patterns, generally ranging between 135-170 hours, with peaks in May 2024 (\~170 hours) and June 2024 (\~169 hours), and a notable dip in August 2024 (\~115 hours). Both groups show declining trends toward the end of the period. Right: A density plot showing the distribution of driver start dates by two groups: 'Paid Same or More' (gray line and shaded area) and 'Paid Less' (red line and shaded area). The x-axis shows start dates from 1.3e18 to 1.8e18 (likely Unix timestamps), while the y-axis shows density from 0 to 5e-18. Both distributions are bell-shaped and largely overlapping, with peaks around 1.45e18. The 'Paid Same or More' group (gray) has a slightly higher peak density of approximately 5e-18, while the 'Paid Less' group (red) has a peak density of about 4.5e-18. The 'Paid Same or More' distribution appears to be slightly more concentrated around the peak, while the 'Paid Less' distribution shows a somewhat wider spread with a longer tail extending to higher start date values. The substantial overlap between the two distributions suggests that driver start dates are relatively similar between the two payment outcome groups.}
\end{figure}

\newpage

\begin{table*}[]
\begin{tabular}{c|ccccccccccc}
\multicolumn{1}{l}{} & Y     & Y-1                  & Y-2                  & Y-3                  & Y-4                  & Y-5                  & \multicolumn{1}{c}{Y-6}    & \multicolumn{1}{c}{Y-7}    & \multicolumn{1}{c}{Y-8}    & \multicolumn{1}{c}{Y-9}    & \multicolumn{1}{c}{Y-10}   \\ \hline
2014                 & 0.873 & \multicolumn{1}{l}{} & \multicolumn{1}{l}{} & \multicolumn{1}{l}{} & \multicolumn{1}{l}{} & \multicolumn{1}{l}{} &                            &                            &                            &                            &                            \\
2015                 & 0.654 & 0.531                & \multicolumn{1}{l}{} & \multicolumn{1}{l}{} & \multicolumn{1}{l}{} & \multicolumn{1}{l}{} &                            &                            &                            &                            &                            \\
2016                 & 0.721 & 0.654                & 0.366                & \multicolumn{1}{l}{} & \multicolumn{1}{l}{} & \multicolumn{1}{l}{} &                            &                            &                            &                            &                            \\
2017                 & 0.772 & 0.747                & 0.532                & 0.028                & \multicolumn{1}{l}{} & \multicolumn{1}{l}{} &                            &                            &                            &                            &                            \\
2018                 & 0.809 & 0.804                & 0.770                & 0.440                & -0.189               & \multicolumn{1}{l}{} &                            &                            &                            &                            &                            \\
2019                 & 0.821 & 0.820                & 0.801                & 0.747                & 0.235                & -0.545               &                            &                            &                            &                            &                            \\
2020                 & 0.830 & 0.829                & 0.823                & 0.794                & 0.697                & -0.161               & \multicolumn{1}{c}{-1.121} &                            &                            &                            &                            \\
2021                 & 0.827 & 0.826                & 0.823                & 0.809                & 0.781                & 0.625                & \multicolumn{1}{c}{-0.707} & \multicolumn{1}{c}{-1.890} &                            &                            &                            \\
2022                 & {\color{red}0.828} & 0.826                & 0.823                & 0.820                & 0.804                & 0.752                & \multicolumn{1}{c}{0.562}  & \multicolumn{1}{c}{-0.899} & \multicolumn{1}{c}{-2.159} &                            &                            \\
2023                 & 0.830 & {\color{red}-3.133}               & -1.269               & -0.794               & -1.033               & -4.499               & \multicolumn{1}{c}{0.690}  & \multicolumn{1}{c}{0.472}  & \multicolumn{1}{c}{-1.028} & \multicolumn{1}{c}{-2.339} &                            \\
2024                 & 0.831 & 0.831                & {\color{red}-3.082}               & -1.243               & -0.773               & -1.010               & \multicolumn{1}{c}{-4.434} & \multicolumn{1}{c}{0.683}  & \multicolumn{1}{c}{0.464}  & \multicolumn{1}{c}{-1.033} & \multicolumn{1}{c}{-2.359}
\end{tabular}
\caption{Predicting price at year $Y$ using years leading upto the year $1, ..., Y-n$}
\label{tab:pred_all}
\end{table*}

\end{document}